# Carbon-Negative Commuting: Integrating Urban Design, Behavior, and Technology for Climate-Positive Mobility

Ebrahim Eslami[*,**,1]
[*] EnviroPilot.ai, Houston, TX
[**] (formerly at) Houston Advanced Research Center (HARC), The Woodlands, TX

**Abstract:** Commuting contributes substantially to urban greenhouse gas emissions and represents a critical focus for climate mitigation efforts. This paper explores the multifaceted nature of commuting-related carbon dioxide emissions by analyzing the influence of urban form, socio-economic attributes, and individual behaviors. It reviews analytical approaches including structural equation modeling, multi-objective optimization, and agent-based simulations that have been employed to understand and mitigate emissions. Building on these insights, the paper develops a conceptual framework for carbon-negative commuting that integrates spatial planning, behavioral interventions, technological innovations, and carbon offsetting strategies. Case studies from diverse global contexts illustrate both the feasibility and challenges of implementing these interventions. The discussion highlights key trade-offs, equity considerations, and governance barriers while identifying co-benefits such as improved public health and urban resilience. The paper concludes by emphasizing the need for interdisciplinary research and adaptive policymaking to operationalize carbon-negative commuting and align urban mobility systems with global decarbonization goals.

**Keywords:** Commuting emissions; carbon-negative mobility; urban form; sustainable transport; behavioral interventions; policy integration; climate mitigation

[1] Email: e.eslami@gmail.com; ebrahim.eslami@enviropilotai.com

## Introduction

Commuting is a fundamental component of daily life in urbanized societies and represents one of the largest contributors to transportation-related greenhouse gas (GHG) emissions. Globally, the transportation sector accounts for approximately 23 to 30 percent of anthropogenic carbon dioxide ($CO_2$) emissions, with commuting trips forming a substantial and dynamic share of this total (Chow, 2016; Aguiléra & Voisin, 2014). As cities expand and sprawl, travel patterns become increasingly complex, creating challenges for mitigating emissions associated with daily mobility. Reducing commuting-related $CO_2$ emissions has thus emerged as a priority for policymakers seeking to meet climate targets such as the Paris Agreement's goal of limiting global warming to 1.5°C.

Urban form exerts a significant influence on commuting emissions by shaping the spatial distribution of residences, employment centers, and transportation networks. Compact, mixed-use urban environments are often associated with shorter travel distances and higher public transport usage, resulting in lower per-capita emissions (Rahman et al., 2023; Sobrino & Arce, 2021, Eslami, 2020). Conversely, urban sprawl tends to increase reliance on private vehicles, exacerbating emissions (Ashik et al., 2023). However, the relationship between urban form and commuting emissions is not universally consistent. Studies in certain contexts have found that high-density areas may exhibit elevated emissions due to traffic congestion, insufficient public transit, or increased vehicle ownership among affluent residents (Aguiléra & Voisin, 2014). These findings underscore the importance of considering both local and regional scales in analyzing the interactions between urban morphology and travel behavior.

Socio-economic factors further mediate the relationship between urban form and commuting emissions. Household income, car ownership, occupational structure, and residential self-selection have all been shown to influence commuting patterns (Jayakrishnan et al., 2023). For example, wealthier households are more likely to own multiple vehicles and undertake longer commutes, even in transit-accessible areas (Sutton-Parker, 2021). Conversely, lower-income individuals may face constraints that limit their mobility options, leading to different patterns of emissions. Behavioral factors, including individual preferences, perceptions of travel time, and willingness to adopt sustainable modes, also play a crucial role (Aziz & Ukkusuri, 2014). The interplay of these socio-economic and behavioral dimensions creates a complex landscape for developing effective emissions reduction strategies.

A variety of analytical approaches have been employed to investigate commuting-related emissions and to design mitigation interventions. Structural equation modeling (SEM) has been used to disentangle direct and indirect effects of urban form and socio-economic attributes on travel behavior and emissions (Ashik et al., 2023). Multi-objective optimization frameworks have sought to balance trade-offs among emissions, travel time, and user preferences (Abdallah et al., 2020). Spatial analyses, including agent-based simulations, have illuminated the heterogeneity of emissions across urban regions and provided insights into the potential of interventions such as congestion pricing, telecommuting, and active transportation promotion (Hatzopoulou et al., 2011). Despite these advances, most studies have focused on either macro-level trends or localized interventions, often neglecting the dynamic interactions among urban systems, individual behaviors, and policy measures.

Recent discourse has introduced the concept of "carbon-negative commuting" as a transformative goal beyond carbon neutrality. In this paradigm, strategies not only aim to reduce emissions but also to offset or sequester $CO_2$ to achieve net-negative outcomes. Achieving this requires integrating multiple levers, including urban planning reforms, technological innovations (such as electric vehicles and renewable energy integration), and behavioral changes (e.g., increased adoption of active and shared mobility modes). Universities, as microcosms of urban systems, have been highlighted as ideal settings for piloting such comprehensive strategies, given their capacity to combine travel demand management with education and innovation (Appleyard et al., 2018; Mathez et al., 2013).

Nevertheless, significant gaps remain in the literature. First, there is a lack of integrative frameworks that couple urban morphology, socio-economic structures, and behavioral interventions within a unified analytical model. Second, the spatial heterogeneity of commuting emissions and the context-specific effectiveness of mitigation measures are often underexplored. Third, while mitigation dominates the research agenda, pathways to carbon-negative commuting remain poorly conceptualized and empirically underdeveloped. Addressing these gaps requires a holistic approach that leverages insights from urban planning, transportation engineering, behavioral science, and environmental modeling.

This paper responds to these challenges by proposing a conceptual framework for achieving carbon-negative commuting. Drawing on a synthesis of existing studies, it examines the drivers of commuting emissions, reviews analytical methods employed to assess and mitigate these emissions, and identifies potential synergies among urban design, socio-economic interventions, and behavioral change. By highlighting pathways to transform commuting from a major source of emissions into a net carbon sink, this work aims to advance the discourse on sustainable urban mobility and offers conceptual and practical guidance for future research and policy development.

## 2. Theoretical Background and Conceptual Foundations

### 2.1 Urban Form and Transportation Emissions

Urban form plays a central role in shaping travel behavior and its environmental impacts, particularly commuting-related $CO_2$ emissions. Compact cities, characterized by high population density, mixed land use, and interconnected street networks, are often associated with shorter travel distances and greater reliance on sustainable transport modes such as walking, cycling, and public transit (Ewing & Cervero, 2010). This urban form is considered conducive to lower per-capita emissions by reducing automobile dependency and fostering efficient land-use patterns. Empirical evidence from European cities supports this assertion. In Barcelona, a study of commuting patterns revealed that policies encouraging densification and improved public transport accessibility significantly reduced emissions from daily travel (Garcia-Sierra & van den Bergh, 2014). Similarly, Newman and Kenworthy's (1999) seminal comparison of global cities highlighted that automobile use and energy consumption per capita are inversely related to urban density.

However, the relationship between urban form and transportation emissions is neither universal nor straightforward. In some contexts, higher densities have been linked to increased congestion and localized air pollution, undermining anticipated environmental benefits (Pam et al., 2019). In rapidly growing megacities of the Global South, such as Lagos and Jakarta, high-density settlements often lack adequate infrastructure and transit systems, resulting in longer travel times and high emissions due to inefficient traffic flow and older vehicle fleets (Rahman et al., 2023). Furthermore, urban compaction may exacerbate the "urban heat island" effect and intensify exposure to pollutants in dense cores, complicating assessments of its net environmental impacts.

Suburbanization and urban sprawl have been widely criticized for fostering car dependency and extending commuting distances, leading to significant increases in $CO_2$ emissions. Aguiléra and Voisin (2014) showed that in French metropolitan regions, residents of low-density suburban areas produced disproportionately higher emissions due to limited access to public transportation and a reliance on private cars for commuting. Similar trends have been observed in North American cities, where postwar suburban development patterns have entrenched automobile-oriented lifestyles (Cervero & Murakami, 2010). These findings underscore the importance of not only promoting density but also ensuring high-quality, accessible, and affordable transit systems to complement urban compaction.

Transit-oriented development (TOD) has emerged as a key planning strategy to address the emissions implications of urban sprawl. By clustering higher-density housing, retail, and employment opportunities around transit nodes, TOD aims to shift travel behavior away from car dependency. Evidence from cities like Vancouver and Curitiba suggests that well-implemented TOD policies can reduce per-capita commuting emissions by encouraging mode shifts to public transit and active transportation (Sobrino & Arce, 2021; Cervero & Murakami, 2010). Yet, the success of TOD is contingent on context-specific factors, including institutional capacity, political will, and socio-economic dynamics, which determine the adoption and effectiveness of such interventions.

**2.2 Socio-economic and Behavioral Determinants**

Beyond urban morphology, socio-economic characteristics exert a powerful influence on commuting patterns and related emissions. Household income, car ownership, employment type, and demographic composition often shape access to mobility options and preferences for certain modes. Higher-income households tend to have greater car ownership and longer commutes, frequently selecting suburban residences that align with lifestyle preferences for space and privacy (Jayakrishnan et al., 2023). In contrast, low-income individuals may live closer to employment centers yet face limited access to reliable public transportation, exposing them to inequitable mobility burdens and higher per-trip emissions in older, less efficient vehicles.

Behavioral dimensions further complicate efforts to reduce commuting emissions. Even when viable alternatives exist, individual preferences for convenience, comfort, and flexibility often sustain private car use. Cultural norms and perceptions of status associated with automobile ownership contribute to this persistence, particularly in emerging economies where rising affluence fuels car acquisition (Sutton-Parker, 2021). Telecommuting and flexible work arrangements, catalyzed by the COVID-19 pandemic, demonstrated substantial potential for emissions reductions, with Sutton-Parker's (2021) study estimating a 97% decrease in commuting emissions during lockdown periods. Yet, the long-term sustainability of remote work and its scalability across different sectors remain uncertain.

The phenomenon of residential self-selection presents a methodological challenge in understanding the impact of urban form on travel behavior. Individuals who prefer sustainable transport modes may actively choose to reside in transit-rich neighborhoods, creating a correlation between urban form and mode choice that does not necessarily imply causation (Handy et al., 2005). Similarly, the adoption of active transportation such as walking and cycling is mediated by factors like perceived safety, weather conditions, and infrastructure quality. Research in Copenhagen and Amsterdam, cities renowned for their cycling cultures, suggests that supportive policies and high-quality infrastructure are critical to sustaining high levels of active commuting and achieving corresponding emissions reductions (Pucher & Buehler, 2008).

## 2.3 Analytical Approaches in Commuting Emissions Studies

A robust body of research has employed diverse analytical methods to examine commuting-related emissions. Structural equation modeling (SEM) has been widely adopted to explore the complex, multi-layered relationships among urban form, socio-economic attributes, and travel behavior. In Dhaka, Ashik et al. (2023) used SEM to demonstrate how built environment variables indirectly influence $CO_2$ emissions by shaping car ownership patterns, revealing the mediating effects of socio-economic factors. These models provide nuanced insights that go beyond simplistic correlations, illuminating the interplay between structural and individual-level determinants of emissions.

Optimization models have also gained prominence in the transportation and environmental planning literature. Multi-objective optimization frameworks allow policymakers to evaluate trade-offs between emissions reduction, travel time, and commuter satisfaction. Abdallah et al. (2020) developed the Business+ Commute Optimization System (B+COS), which identified Pareto-optimal solutions for businesses aiming to minimize employee commute times and emissions. Such approaches provide actionable strategies for decision-makers balancing environmental goals with operational efficiency.

Agent-based modeling and activity-based travel demand simulations offer further analytical sophistication by capturing heterogeneity in individual behaviors and interactions within transportation systems. Hatzopoulou et al. (2011) integrated emission modeling with an activity-based framework for the Greater Toronto Area, enabling assessment of how land-use changes and transportation policies affect emissions and population exposure to air pollution. These methods, while powerful, require significant data inputs and computational resources, limiting their application in data-scarce regions.

## 2.4 Defining Carbon-negative Commuting

The concept of carbon-negative commuting represents a paradigmatic shift in sustainable mobility thinking. Rather than focusing solely on reducing emissions, carbon-negative commuting aims to create net-negative outcomes by combining aggressive mitigation strategies with offsetting and sequestration measures. This conceptualization aligns with broader climate goals articulated in frameworks such as net-zero emissions and climate-positive urban development.

Achieving carbon-negative commuting entails integrating multiple strategies. Promoting active transportation and public transit can directly reduce emissions, while technological innovations such as widespread electrification of transport systems powered by renewable energy further diminish carbon footprints. Urban greening initiatives, such as planting vegetation along transport corridors and incorporating green roofs on transit infrastructure, provide co-benefits of carbon sequestration and air quality improvement (Appleyard et al., 2018). Additionally, institutional policies supporting telecommuting, dynamic congestion pricing, and shared mobility services can amplify these effects.

Despite its promise, operationalizing carbon-negative commuting faces significant challenges. Limited empirical evidence, infrastructural constraints, and behavioral inertia hinder the translation of conceptual frameworks into actionable policies. Moreover, socio-economic disparities may exacerbate inequities if interventions disproportionately benefit affluent populations with access to sustainable transport options while marginalizing low-income commuters.

## 3. Methodological Synthesis: Framework for Carbon-Negative Commuting

Transforming commuting from a significant source of greenhouse gas emissions into a potential net carbon sink requires a comprehensive and integrative methodological approach that accounts for the multidimensional drivers of travel behavior and emissions. A synthesis of existing analytical perspectives suggests that urban morphology, socio-economic dynamics, and behavioral factors interact within complex systems that both constrain and enable transitions toward carbon-negative commuting. This section proposes a conceptual framework that integrates these dimensions, offering a pathway to operationalize and evaluate carbon-negative strategies within urban transportation systems.

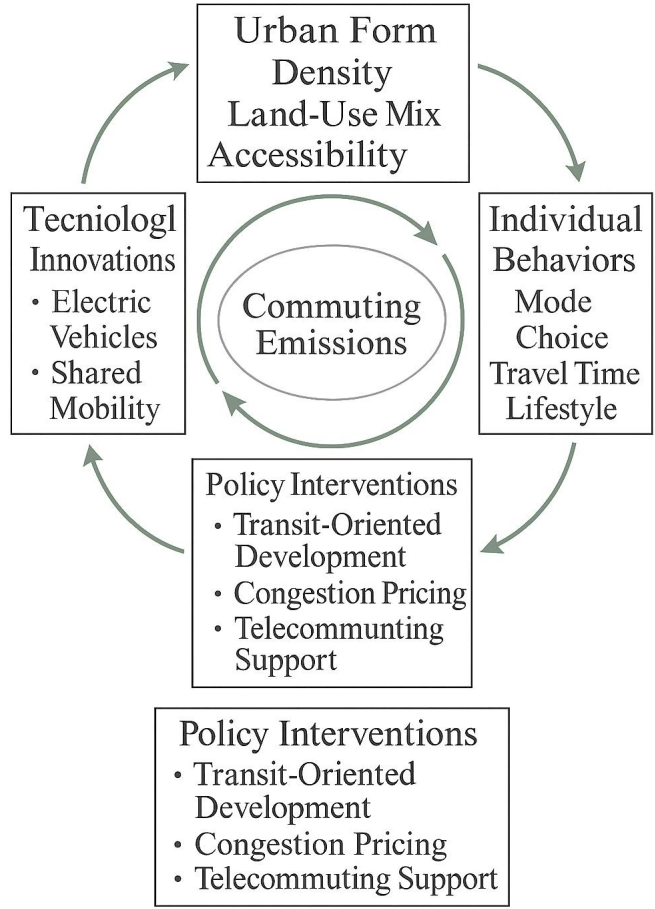

Figure 1. Conceptual framework that synthesize the relationships among urban form, socio-economic factors, behavior, technological interventions, and policy levers that shape commuting emissions.

## 3.1 Integrating Urban Form, Socio-Economic Factors, and Behavior

At the core of the proposed framework is the recognition that commuting emissions arise from the intersection of spatial structures and human decision-making processes. Urban form establishes the physical context within which commuting behaviors occur, influencing trip distances, modal choices, and accessibility to employment and services. Compact, polycentric, and transit-oriented urban forms are associated with reduced per-capita emissions, as highlighted in Section 2, but these effects are mediated by socio-economic conditions such as household income, car ownership, and occupational patterns (Rahman et al., 2023; Chow, 2016).

Empirical examples reinforce these theoretical insights. In Stockholm, the integration of dense urban neighborhoods with high-quality public transportation reduced private car dependency, leading to a 25 percent decrease in per-capita commuting emissions over two decades (Svensson & Johansson, 2018). In contrast, sprawling metropolitan areas such as Houston exhibit persistently high emissions due to low-density development and limited transit coverage, underscoring the importance of urban form in shaping emissions outcomes (Newman & Kenworthy, 1999).

Socio-economic attributes shape mobility preferences and constraints, determining the feasibility and attractiveness of low-carbon commuting options. For instance, high-income households in London's metropolitan region, despite having access to extensive transit infrastructure, maintain significant levels of car use due to lifestyle preferences, while lower-income groups rely heavily on buses and walking, albeit often in suboptimal conditions (Banister, 2011). Behavioral factors, including habitual car use, perceptions of convenience and safety, and responsiveness to environmental information, further condition the effectiveness of interventions targeting commuting emissions (Aziz & Ukkusuri, 2014).

The framework posits a dynamic interaction between these dimensions, where changes in one domain can produce feedback effects in others. For instance, investments in public transport infrastructure can alter travel behaviors and socio-economic accessibility, while policy measures such as parking pricing and congestion charges can reshape residential location choices and modal preferences over time. Recognizing these feedback loops is essential for designing effective carbon-negative commuting strategies.

### 3.2 Analytical Model Components

The conceptual framework integrates multiple methodological components to capture the complexity of commuting emissions and identify intervention points:

**Spatial Analysis and Urban Morphology:** Spatially explicit models analyze how land use, employment distributions, and transportation networks influence commuting patterns. Geographical information systems (GIS) and spatial regression techniques can identify emission hotspots and evaluate the potential of urban design interventions, such as densification and transit-oriented development, to reduce carbon footprints (Cervero & Murakami, 2010). A recent study in Tel Aviv mapped intra-city commuting emissions at a neighborhood level, revealing that targeted densification near rail nodes could cut emissions by up to 40 percent (Kissinger & Reznik, 2019).

**Behavioral Economics and Travel Demand Modeling:** Behavioral models, including agent-based and activity-based simulations, provide insights into how individual commuters respond to policies such as dynamic pricing, telecommuting incentives, and improvements in active transportation infrastructure. These approaches enable scenario testing under diverse socio-economic and cultural contexts, allowing policymakers to anticipate unintended consequences and equity implications (Hatzopoulou et al., 2011). In Singapore, an agent-based model evaluating road pricing strategies demonstrated that congestion pricing could reduce peak-hour vehicle kilometers traveled (VKT) by 18 percent while encouraging a modal shift to rail transit (Zhang et al., 2020).

**Emissions Accounting and Lifecycle Assessment:** Robust accounting methods quantify direct and indirect emissions from commuting activities. Lifecycle assessment frameworks incorporate upstream emissions from vehicle manufacturing, infrastructure development, and energy production, providing a holistic view of transportation systems' carbon footprints. A study of employee commutes across three large healthcare systems in Cleveland, Ohio, revealed that employee travel accounted for 36,000 metric tons of $CO_2$ annually, highlighting the significance of integrating commuting patterns into institutional decarbonization plans (Jayakrishnan et al., 2023).

**Multi-Objective Optimization and Trade-off Analysis:** Multi-objective optimization frameworks enable simultaneous consideration of competing priorities such as emissions reduction, travel time minimization, and cost efficiency. Pareto frontier analyses, as demonstrated by Abdallah et al. (2020), help identify policy configurations that balance environmental goals with user satisfaction, providing decision-makers with a menu of optimal solutions. In a case study from Barcelona, policy scenarios combining public transport investment and active commuting incentives achieved emissions reductions of up to 35 percent while minimizing travel time increases (Garcia-Sierra & van den Bergh, 2014).

### 3.3 Potential Pathways and Levers for Intervention

The framework identifies several synergistic pathways to achieve carbon-negative commuting:

**Promoting Active Transportation and Public Transit:** Investments in pedestrian and cycling infrastructure, combined with high-frequency, affordable public transit, can induce modal shifts away from private car use. Studies in Copenhagen and Amsterdam underscore the potential of integrated active transportation networks to significantly reduce commuting emissions while delivering co-benefits for public health and urban livability (Pucher & Buehler, 2008). In Bogotá, Colombia, the implementation of a city-wide bike-sharing system coupled with new cycle tracks increased cycling mode share from 0.6 to 3.2 percent within five years, reducing estimated annual commuting emissions by over 50,000 metric tons of $CO_2$ (Cervero et al., 2009).

**Leveraging Technological Innovations:** The electrification of vehicle fleets, coupled with decarbonized energy grids, represents a critical technological pathway. Shared mobility platforms and autonomous vehicle technologies offer opportunities to increase vehicle occupancy rates and optimize routing, further mitigating emissions. A study of private autonomous vehicles (PAVs) in Toronto found that, if integrated with public transit and subject to appropriate pricing policies, PAVs could reduce household vehicle ownership and emissions by 12 percent over a decade (Saleh & Hatzopoulou, 2020).

**Implementing Behavioral and Policy Interventions:** Policies such as congestion pricing, parking management, and employer-based telecommuting incentives can influence commuter choices and reduce peak-hour travel demand. Behavioral nudges, including information campaigns and gamification strategies, can reinforce sustainable mobility practices. In Japan, traveler feedback programs providing real-time emissions data to commuters achieved a measurable reduction in vehicle use and induced a 10 percent increase in public transport ridership (Taniguchi et al., 2018).

**Offsetting and Sequestration Measures:** To achieve net-negative outcomes, direct emissions reductions must be complemented by carbon sequestration strategies. Urban greening initiatives, such as planting trees along transport corridors and incorporating green infrastructure into transit systems, can capture $CO_2$ and provide ancillary benefits such as improved air quality and urban cooling (Appleyard et al., 2018). In Melbourne, an urban forest strategy integrated with transportation corridors was estimated to sequester 1.2 million tons of $CO_2$ over 20 years while reducing commuter exposure to heat stress (City of Melbourne, 2016).

The framework underscores the necessity of multi-level governance and cross-sectoral collaboration to coordinate these interventions. Engaging stakeholders, including local governments, employers, community organizations, and commuters themselves, is vital for building the institutional capacity and social support required to operationalize carbon-negative commuting.

## 4. Policy Implications and Strategies

Realizing the vision of carbon-negative commuting requires a multidimensional policy approach that addresses the interconnected nature of urban form, socio-economic dynamics, technological innovation, and individual behavior. Policymakers at municipal, regional, and national levels are increasingly challenged to develop strategies that simultaneously reduce commuting-related greenhouse gas emissions and promote equity, resilience, and livability in urban systems. Drawing on the conceptual framework outlined in the previous section, this discussion identifies key policy domains and strategic levers for advancing carbon-negative commuting, while critically reflecting on barriers and enabling conditions for implementation.

Table 1. Policy-Strategy Matrix: summarizing interventions by domain and expected impact

| Domain | Intervention | Expected Impact | Challenges |
|---|---|---|---|
| Urban Planning | Transit-Oriented Development (TOD) | Reduced car dependency, emissions | Gentrification, land-use constraints |
| Behavioral Interventions | Congestion Pricing | Modal shift to public transport | Public resistance, equity concerns |
| Technological Innovations | Fleet Electrification | Lower operational emissions | Grid decarbonization needed |
| Offsetting Mechanisms | Urban Greening | $CO_2$ sequestration, heat mitigation | Maintenance, limited urban space |

### 4.1 Urban Planning for Sustainable Mobility

Urban planning policies are foundational in shaping commuting patterns and, by extension, their environmental impacts. Compact, polycentric urban forms and transit-oriented development (TOD) are consistently associated with reduced reliance on private vehicles and lower per-capita emissions (Cervero & Murakami, 2010, Xu et al., 2025). Policies that encourage mixed-use zoning, higher residential densities near employment centers, and integration of affordable housing with public transit corridors are critical to enabling sustainable commuting.

International examples illustrate the efficacy of such interventions. In Freiburg, Germany, the Vauban neighborhood was designed as a car-free district with dense, mixed-use development and high-quality cycling and transit infrastructure. As a result, 70 percent of residents commute via walking, cycling, or public transport, achieving per-capita emissions 60 percent lower than citywide averages (Scheurer & Newman, 2009). Similarly, Curitiba, Brazil, has demonstrated how well-planned bus rapid transit (BRT) systems can catalyze compact urban growth and shift commuting patterns toward low-carbon modes (Cervero et al., 2009).

For cities with entrenched sprawl, retrofitting suburban environments with localized employment hubs, micro-mobility networks, and enhanced transit services is essential. These interventions can reduce long-distance commuting while creating vibrant, low-carbon neighborhoods.

## 4.2 Behavioral Interventions and Demand Management

Behavioral change is a critical complement to structural interventions. Travel demand management (TDM) policies aim to influence commuter choices by altering the relative costs and benefits of different modes. Congestion pricing, parking reform, and employer-based programs are among the most promising tools for reducing car dependency.

Singapore's congestion pricing scheme, first implemented in the 1970s, remains a global benchmark for managing peak-hour traffic and incentivizing modal shifts to public transit (Zhang et al., 2020). More recently, London's congestion charge has reduced vehicle kilometers traveled (VKT) within the charging zone by 20 percent and cut $CO_2$ emissions by 15 percent over a decade (Beevers & Carslaw, 2005).

Employer-based initiatives, such as subsidized transit passes, carpool matching services, and flexible work schedules, can further enhance commuting sustainability. The University of California, Davis, achieved a 30 percent reduction in single-occupancy vehicle commuting through a suite of such programs, demonstrating the potential of institutional actors to drive behavioral change (Shoup, 2005).

Public education campaigns and social marketing strategies also play a vital role. Providing commuters with real-time information on emissions, costs, and travel times can encourage informed decision-making. In Tokyo, traveler feedback programs resulted in measurable reductions in private car use and increased public transport ridership (Taniguchi et al., 2018).

## 4.3 Technological Innovations and Offsetting Mechanisms

Technological innovation is indispensable in the transition to carbon-negative commuting. Electrification of transport systems, when paired with renewable energy generation, can significantly reduce direct emissions. Cities such as Oslo and Shenzhen have pioneered the electrification of municipal bus fleets, achieving dramatic reductions in operational emissions and demonstrating scalable models for other urban contexts (Huang et al., 2021).

Shared mobility platforms and autonomous vehicle (AV) technologies offer further opportunities to reduce emissions, provided they are carefully managed to avoid rebound effects such as increased vehicle miles traveled. Policy frameworks that prioritize shared, electric, and autonomous mobility (SEAM) can align technological advances with environmental goals.

Beyond direct reductions, urban greening and carbon sequestration initiatives provide critical offsetting mechanisms. Tree planting programs, green roofs, and urban forest strategies not only capture $CO_2$ but also improve air quality, mitigate urban heat islands, and enhance aesthetic appeal. Melbourne's Urban Forest Strategy, for example, is projected to sequester over one million tons of $CO_2$ while reducing commuter exposure to heat stress (City of Melbourne, 2016).

**4.4 Implementation Challenges and Political Economy Dynamics**

Despite the growing body of evidence on effective interventions for reducing commuting emissions, implementing carbon-negative commuting strategies faces substantial barriers rooted in institutional inertia, political economy dynamics, and socio-cultural resistance. Understanding these challenges is critical to designing policies that are not only technically sound but also socially and politically viable.

One of the most significant barriers is the entrenched dominance of automobile-oriented planning paradigms. Decades of investment in highway infrastructure and suburban development have created powerful vested interests, including automotive manufacturers, fuel providers, and real estate developers, whose business models are tied to high levels of car dependency. In many contexts, these actors exert considerable influence over urban policy agendas, often resisting reforms such as road pricing or reductions in parking supply (Shoup, 2005). Overcoming these structural lock-ins requires strong political leadership and coalitions of supportive stakeholders, including environmental NGOs, public health advocates, and progressive business interests.

Public resistance to demand management policies, such as congestion pricing and parking fees, represents another formidable challenge. Such measures are frequently perceived as regressive or punitive, particularly in societies where automobile ownership is closely associated with personal freedom and social status. The political backlash against Stockholm's congestion charge in the early 2000s illustrates how opposition can initially derail ambitious policies. However, once implemented and accompanied by tangible improvements in air quality and public transit services, public support grew significantly, leading to the charge's permanent adoption (Eliasson, 2014). This case underscores the importance of framing policies around co-benefits and ensuring transparent use of revenue to enhance alternative transport modes.

Equity concerns are also central in the political economy of sustainable commuting. Without deliberate attention to distributive impacts, interventions such as TOD and electrification risk exacerbating socio-spatial inequalities. For instance, TOD projects in U.S. cities have sometimes triggered gentrification, displacing low-income residents from transit-accessible neighborhoods (Jones & Ley, 2016). Policies that integrate affordable housing provisions and prioritize underserved communities in transit investments are essential to avoid reproducing patterns of exclusion.

Institutional fragmentation further complicates implementation. Commuting systems span multiple jurisdictions, with land use, transportation, and environmental policies often governed by separate agencies with competing mandates and limited coordination. Successful models, such as Copenhagen's "Finger Plan" integrating land use and rail corridors, demonstrate the benefits of regional governance frameworks that align spatial planning and mobility goals (Knowles, 2012). Building similar governance capacity in other contexts remains a critical priority.

Finally, global economic and political trends influence the feasibility of carbon-negative strategies. Energy price volatility, fiscal constraints, and shifts in public priorities during crises (such as pandemics or geopolitical conflicts) can derail long-term investments in sustainable transport. To achieve lasting transformation, policymakers must therefore design adaptive, resilient policies that can sustain momentum amid uncertainty, while building broad-based societal coalitions to support transformative change.

While the preceding sections have identified a range of interventions and governance strategies to enable carbon-negative commuting, their practical implementation requires navigating a complex terrain of trade-offs, synergies, and uncertainties. Translating conceptual frameworks into actionable policies necessitates not only technical and financial resources but also a critical awareness of context-specific challenges and opportunities. The following section reflects on these dimensions, synthesizing key insights and highlighting research and policy directions to advance the discourse on carbon-negative commuting.

Table 2. An overview of cities implementing carbon-negative commuting components.

| City | Key Intervention | Results | Lessons Learned |
|---|---|---|---|
| Copenhagen | Active transport infrastructure | 62% bike mode share, emissions reductions | Consistent investment and cultural support |
| Singapore | Congestion pricing | 18% decrease in peak-hour VKT | Revenue reinvestment in public transport |
| Freiburg | Car-free district planning (Vauban) | 70% non-motorized commuting | Integration of housing and transport planning |
| Bogotá | BRT and bike-sharing | Significant increase in active commuting modes | Political will critical for scaling efforts |

## 5. Discussion

Achieving carbon-negative commuting represents an ambitious yet essential paradigm shift in urban mobility, requiring systemic interventions across technological, behavioral, and governance domains to align with global climate mitigation targets. Building on the conceptual framework and policy interventions outlined in earlier sections, this discussion critically examines the trade-offs and co-benefits of implementing carbon-negative commuting strategies, identifies key challenges in operationalizing such a transition, and suggests future research directions to advance both theoretical and practical understanding.

Table 3. The political economy and institutional factors influencing success.

| Barrier | Potential Enabler |
|---|---|
| Institutional fragmentation | Metropolitan governance frameworks |
| Public resistance to pricing mechanisms | Transparent revenue use and communication efforts |
| High upfront infrastructure costs | Green bonds and climate finance mechanisms |
| Equity concerns (e.g., gentrification) | Affordable housing mandates in TOD zones |

## 5.1 Trade-offs and Co-Benefits

Efforts to achieve carbon-negative commuting inevitably involve navigating trade-offs among environmental, social, and economic objectives. For example, densification policies aimed at reducing commuting distances and encouraging active transportation can inadvertently lead to rising property values and gentrification pressures, displacing low-income residents from transit-accessible neighborhoods (Jones & Ley, 2016). Such socio-spatial inequities risk undermining the equity goals that are increasingly central to sustainable urbanism. Conversely, carefully designed inclusionary zoning policies and investments in affordable housing near transit hubs can help balance environmental objectives with social justice considerations.

Technological innovations such as electrification and shared autonomous vehicles also present complex trade-offs. While electrification reduces operational emissions, lifecycle emissions associated with battery production and disposal remain significant, particularly in regions reliant on fossil fuel-based electricity grids (Hawkins et al., 2013). Shared autonomous vehicle systems, if poorly regulated, could induce greater travel demand and congestion, a phenomenon known as the "rebound effect" (Zhang et al., 2020). Policymakers must therefore implement complementary measures such as dynamic pricing and prioritization of high-occupancy modes to mitigate these risks.

Despite these challenges, co-benefits of carbon-negative commuting strategies offer powerful justifications for action. Investments in active transportation infrastructure improve public health outcomes by promoting physical activity and reducing exposure to air pollution (Pucher & Buehler, 2008). Urban greening initiatives not only sequester carbon but also enhance climate resilience by mitigating heat island effects and managing stormwater runoff (City of Melbourne, 2016). These synergies highlight the potential of integrated planning approaches to deliver multiple societal benefits beyond carbon mitigation.

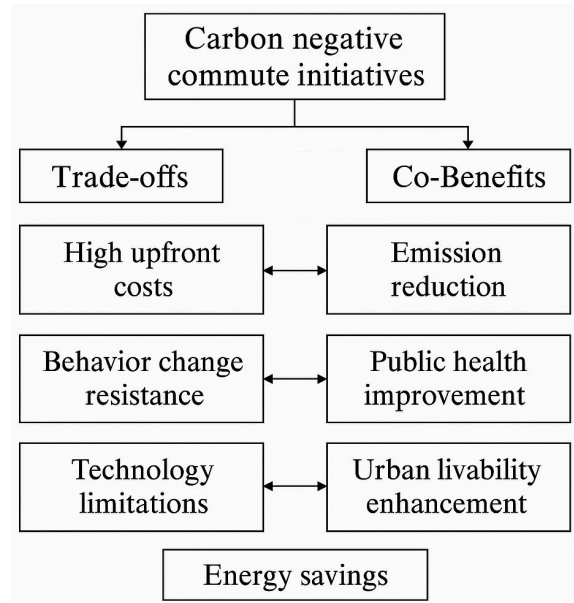

Figure 2. Trade-offs and Co-Benefits flowchart.

### 5.2 Challenges in Operationalizing Carbon-Negative Commuting

Operationalizing carbon-negative commuting requires addressing a series of interrelated challenges at institutional, technical, and socio-cultural levels. As noted in Section 4, institutional fragmentation often impedes coordinated action across land use, transportation, and environmental agencies. In many metropolitan regions, overlapping jurisdictions and competing mandates create policy silos that hinder the implementation of holistic commuting strategies (Banister, 2011). Establishing metropolitan governance structures with the authority to align spatial planning and mobility goals is essential for overcoming these barriers.

Technical and financial constraints pose additional obstacles. The transition to carbon-negative commuting demands substantial upfront investments in public transport infrastructure, electric vehicle charging networks, and urban greening initiatives. Securing sustainable financing mechanisms, such as congestion pricing revenues or green bonds, is critical to bridging these funding gaps (Eliasson, 2014). In lower-income countries, where fiscal resources and institutional capacity are limited, international cooperation and climate finance mechanisms may be necessary to support transformative investments.

Socio-cultural resistance further complicates implementation. In many contexts, automobile ownership is deeply embedded in social norms and associated with status, convenience, and autonomy. Demand management measures such as congestion pricing and parking restrictions often encounter public opposition, particularly when perceived as regressive or punitive (Shoup, 2005). To build political legitimacy, policies must be designed with transparency and fairness, accompanied by robust public engagement and reinvestment of revenues in high-quality alternative transport options.

## 5.3 Future Research Directions

To advance the carbon-negative commuting agenda, future research must address critical gaps in knowledge and practice. First, empirical studies are needed to evaluate the long-term effectiveness of integrated urban planning, technological, and behavioral interventions across diverse socio-economic and cultural contexts. While case studies from cities such as Copenhagen, Freiburg, and Curitiba provide valuable insights, evidence from rapidly urbanizing regions in Asia, Africa, and Latin America remains limited.

Second, methodological innovations are required to capture the dynamic interactions and feedback loops between urban form, socio-economic structures, and individual behaviors. Hybrid modeling approaches that combine structural equation modeling (SEM), agent-based simulations, and lifecycle assessment can provide more robust tools for assessing intervention impacts and unintended consequences (Ashik et al., 2023; Hatzopoulou et al., 2011).

Third, greater attention must be paid to equity implications. Research should explore how carbon-negative commuting strategies can be designed to benefit marginalized populations and avoid exacerbating existing inequalities. Participatory approaches that engage communities in the co-design of interventions hold promise for enhancing both effectiveness and legitimacy.

Finally, interdisciplinary collaboration between urban planners, engineers, behavioral scientists, and policymakers is essential to bridge the gap between theory and practice. Integrating climate mitigation goals with broader sustainability and livability agendas can help mainstream carbon-negative commuting as a central component of urban development strategies.

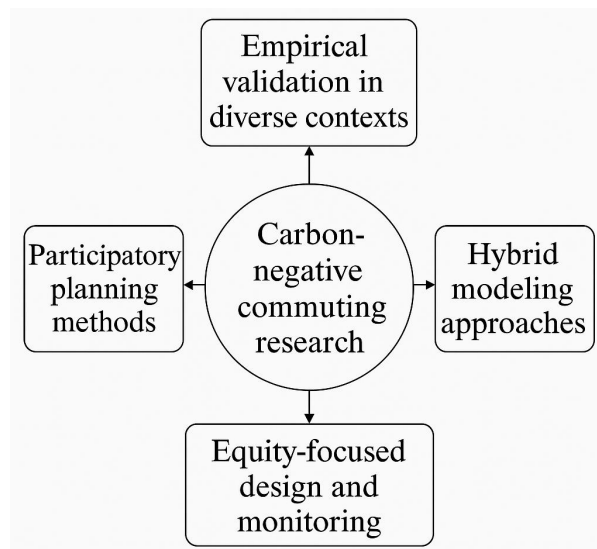

Figure 3. Research gaps and future direction

## 6. Conclusion

Commuting represents a critical arena for addressing the intertwined challenges of urban sustainability and climate change mitigation. As this paper has demonstrated, commuting-related greenhouse gas emissions are shaped by a complex interplay of urban form, socio-economic structures, and individual behaviors, mediated further by technological and policy contexts. Achieving carbon-negative commuting requires moving beyond incremental mitigation strategies to embrace transformative, systemic approaches that integrate spatial planning, demand management, technological innovation, and carbon offsetting.

The conceptual framework developed in this paper offers a multidimensional perspective on the pathways and levers available to transition commuting systems toward net-negative emissions. It highlights how interventions at the urban scale, such as compact, polycentric development and transit-oriented planning, can be complemented by behavioral change policies, technological advances like electrification and shared mobility, and offsetting measures through urban greening and sequestration. Case studies from cities including Copenhagen, Freiburg, Curitiba, and Singapore illustrate the feasibility of these strategies, while also underscoring the importance of context-sensitive design and governance.

Yet, significant challenges remain. Political economy dynamics, institutional fragmentation, and socio-cultural resistance often impede the implementation of ambitious commuting reforms. Addressing these barriers requires strong political leadership, inclusive governance structures, and mechanisms to ensure equity and public support. Policies must be designed to deliver co-benefits such as improved public health, enhanced air quality, and greater urban resilience, thereby strengthening societal buy-in and long-term sustainability.

Future research should focus on empirically testing the integrated strategies proposed in this framework across diverse urban contexts, particularly in rapidly urbanizing regions of the Global South where commuting patterns and infrastructural challenges differ markedly from high-income settings. Interdisciplinary collaboration will be essential to develop robust modeling tools, monitor the social and environmental impacts of interventions, and inform adaptive governance.

As cities confront the urgent need to decarbonize, commuting offers both a significant challenge and a transformative opportunity. By embedding carbon-negative goals within broader urban development agendas, policymakers and researchers can contribute to building cities that are not only low-carbon but also more equitable, healthy, and resilient. The transition to carbon-negative commuting is ambitious, but with coordinated action and innovation, it is a vision within reach.

**Acknowledgment:**

This work was carried out independently by the author. A portion of the effort benefited from professional development support provided by the Houston Advanced Research Center (HARC). The views and conclusions expressed here are solely those of the author and do not necessarily reflect those of HARC.

# References

Abdallah, M., Tawfik, A. M., Monghasemi, S., Clevenger, C. M., & Adame, B. A. (2020). Developing commute optimization system to minimize negative environmental impacts and time of business commuters. *International Journal of Sustainable Transportation, 14*(2), 101–119. https://doi.org/10.1080/15568318.2018.1531184

Aguiléra, A., & Voisin, M. (2014). Urban form, commuting patterns and $CO_2$ emissions: What differences between the municipality's residents and its jobs? *Transportation Research Part A: Policy and Practice, 69*, 243–251. https://doi.org/10.1016/j.tra.2014.07.012

Appleyard, B., Frost, A. R., Cordova, E., & McKinstry, J. (2018). Pathways toward zero-carbon campus commuting: Innovative approaches in measuring, understanding, and reducing greenhouse gas emissions. *Transportation Research Record, 2672*(24), 87–97. https://doi.org/10.1177/0361198118798238

Ashik, F. R., Rahman, M. H., Antipova, A., & Zafri, N. M. (2023). Analyzing the impact of the built environment on commuting-related carbon dioxide emissions. *International Journal of Sustainable Transportation, 17*(3), 258–272. https://doi.org/10.1080/15568318.2022.2031356

Banister, D. (2011). The trilogy of distance, speed and time. *Journal of Transport Geography, 19*(4), 950–959. https://doi.org/10.1016/j.jtrangeo.2010.12.004

Beevers, S. D., & Carslaw, D. C. (2005). The impact of congestion charging on vehicle emissions in London. *Atmospheric Environment, 39*(1), 1–5. https://doi.org/10.1016/j.atmosenv.2004.10.001

Cervero, R., & Murakami, J. (2010). Effects of built environments on vehicle miles traveled: Evidence from 370 US urbanized areas. *Environment and Planning A, 42*(2), 400–418. https://doi.org/10.1068/a4236

Cervero, R., Sarmiento, O., Jacoby, E., Gomez, L. F., & Neiman, A. (2009). Influences of built environments on walking and cycling: Lessons from Bogotá. *International Journal of Sustainable Transportation, 3*(4), 203–226. https://doi.org/10.1080/15568310802178314

Chow, A. S. Y. (2016). Spatial-modal scenarios of greenhouse gas emissions from commuting in Hong Kong. *Journal of Transport Geography, 54*, 205–213. https://doi.org/10.1016/j.jtrangeo.2016.06.001

City of Melbourne. (2016). *Urban Forest Strategy: Making a great city greener 2012–2032*. Melbourne: City of Melbourne.

Eliasson, J. (2014). The Stockholm congestion charges: An overview. *Centre for Transport Studies Stockholm, CTS Working Paper 2014:7.*

Eslami, E. (2020). Applications of Deep Learning in Atmospheric Sciences: Air Quality Forecasting, Post-Processing, and Hurricane Tracking (Doctoral dissertation, University of Houston).

Hatzopoulou, M., Hao, J. Y., & Miller, E. J. (2011). Simulating the impacts of household travel on greenhouse gas emissions, urban air quality, and population exposure. *Transportation, 38*(6), 871–887. https://doi.org/10.1007/s11116-011-9362-9

Hawkins, T. R., Singh, B., Majeau-Bettez, G., & Strømman, A. H. (2013). Comparative environmental life cycle assessment of conventional and electric vehicles. *Journal of Industrial Ecology, 17*(1), 53–64. https://doi.org/10.1111/j.1530-9290.2012.00532.x

Huang, Y., Yu, S., & Liu, Y. (2021). Towards zero-emission public transport: A case study of Shenzhen's electric bus program. *Energy Policy, 149*, 112066. https://doi.org/10.1016/j.enpol.2020.112066

Jayakrishnan, T., Gordon, I. O., O'Keeffe, S., Singh, M. K., & Sehgal, A. R. (2023). The carbon footprint of health system employee commutes. *The Journal of Climate Change and Health, 11*, 100216. https://doi.org/10.1016/j.joclim.2023.100216

Jones, C. E., & Ley, D. (2016). Transit-oriented development and gentrification along Metro Vancouver's low-income SkyTrain corridor. *The Canadian Geographer, 60*(1), 9–22. https://doi.org/10.1111/cag.12256

Kissinger, M., & Reznik, A. (2019). Detailed urban analysis of commute-related GHG emissions to guide urban mitigation measures. *Environmental Impact Assessment Review, 76*, 26–35. https://doi.org/10.1016/j.eiar.2019.01.003

Knowles, R. D. (2012). Transit-oriented development in Copenhagen, Denmark: From the Finger Plan to Ørestad. *Journal of Transport Geography, 22*, 251–261. https://doi.org/10.1016/j.jtrangeo.2012.01.009

Pan, S., Roy, A., Choi, Y., Eslami, E., Thomas, S., Jiang, X., & Gao, H. O. (2019). Potential impacts of electric vehicles on air quality and health endpoints in the Greater Houston Area in 2040. Atmospheric Environment, 207, 38-51.

Pucher, J., & Buehler, R. (2008). Making cycling irresistible: Lessons from the Netherlands, Denmark and Germany. *Transport Reviews, 28*(4), 495–528. https://doi.org/10.1080/01441640701806612

Scheurer, J., & Newman, P. (2009). Vauban: A European model bridging the green and brown agendas. *Australian Planner, 46*(2), 38–47. https://doi.org/10.1080/07293682.2009.9982211

Shoup, D. (2005). *The high cost of free parking*. Chicago: Planners Press.

Taniguchi, A., Suzuki, H., & Fujii, S. (2018). Mobility management in Japan: Its development and meta-analysis of travel feedback programs. *Transport Policy, 63*, 127–135. https://doi.org/10.1016/j.tranpol.2017.12.016

Xu, Y., Willis, S., Dallmann, A., Cai, H., Hu, L., Zou, L., ... & Ye, X. (2025). Urban Climate Adaptation and REsilience (U-CARE) in Texas: Insights from Interdisciplinary Perspectives. Applied Spatial Analysis and Policy, 18(2), 53.

Zhang, W., Guhathakurta, S., Fang, J., & Zhang, G. (2020). Exploring the impact of shared autonomous vehicles on urban parking demand: An agent-based simulation approach. *Sustainable Cities and Society, 53*, 101967. https://doi.org/10.1016/j.scs.2019.101967